\documentclass[prx,aps,twocolumn,notitlepage,superscriptaddress,showpacs,nofootinbib]{revtex4-2}
\usepackage{etoolbox}
\newtoggle{supp}
\togglefalse{supp}  
\usepackage{enumerate,appendix}
\usepackage{qcircuit}
\usepackage{amsmath, amsthm, amssymb,commath}
\usepackage{color,calc,graphicx}
\usepackage[usenames,dvipsnames,svgnames,table,cmyk,hyperref]{xcolor}
\usepackage[colorlinks]{hyperref}
\usepackage{optidef}
\hypersetup{
	colorlinks = true,
	urlcolor = {blue},
	citecolor = {blue},
	linkcolor= {blue}
}

\usepackage{graphicx}
\usepackage{amsmath}
\usepackage{latexsym}

\usepackage[charter,cal=cmcal,sfscaled=false]{mathdesign}
\usepackage{booktabs}
\title{Bibliography management: \texttt{natbib} package}
\usepackage{graphicx, caption, subcaption}
\graphicspath{ {./images/} }
\usepackage{natbib}
\usepackage{multirow} 
\usepackage{dcolumn}
\usepackage{mathrsfs}
\usepackage{csvsimple-l3}

\def \be {\begin{equation}}
\def \ee {\end{equation}}

\newcommand{\tr}{\mathrm{Tr}}
\newcommand{\Tr}{\mathrm{Tr}}

\newcommand{\ket}[1]{|#1\rangle}
\newcommand{\bra}[1]{\langle#1|}

\def\>{\rangle}
\def\<{\langle}

\begin{document}

\title{Magic state distillation with permutation-invariant codes and a two-qubit example}

\author{Heather Leitch}
\email{h.leitch@sheffield.ac.uk}
\affiliation{School of Mathematical and Physical Sciences, University of Sheffield, Sheffield, S3 7RH, United Kingdom}
\author{ Yingkai Ouyang}
\email{y.ouyang@sheffield.ac.uk}
\affiliation{School of Mathematical and Physical Sciences, University of Sheffield, Sheffield, S3 7RH, United Kingdom}

\begin{abstract} 
\centering
Magic states, by allowing non-Clifford gates through gate teleportation, are important building blocks of fault-tolerant quantum computation. Magic state distillation protocols aim to create clean copies of magic states from many noisier copies. However, the prevailing protocols require substantial qubit overhead. We present a distillation protocol based on permutation-invariant gnu codes, as small as two qubits. The two-qubit protocol achieves a 0.5 error threshold and 1/2 distillation rate, surpassing prior schemes for comparable codes. Our protocol furthermore distils magic states with arbitrary magic by varying the position of the ideal input states on the Bloch sphere. We achieve this by departing from the usual magic state distillation formalism, allowing the use of non-Clifford gates in the distillation protocol, and allowing the form of the output state to differ from the input state. Our protocol is compatible for use in tandem with existing magic state distillation protocols to enhance their performance.
\end{abstract}

\maketitle

\section{Introduction}

\begin{figure*}[htbp]
\centering
\makebox[\textwidth][c]{%
    \includegraphics[width=1.1\textwidth]{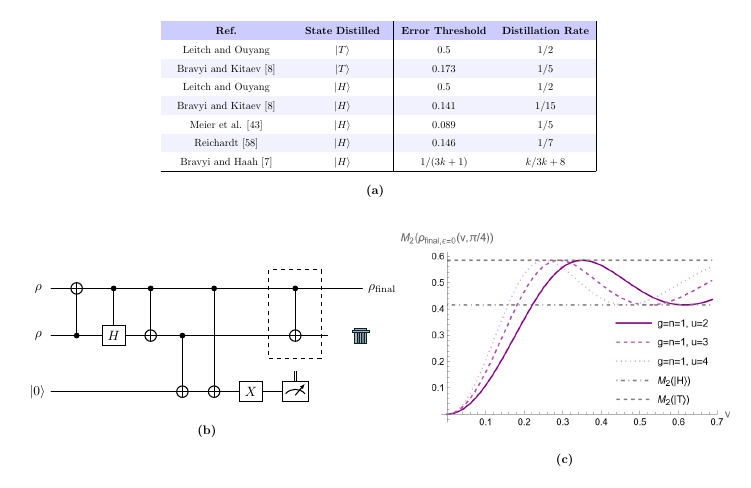}%
}
\caption{(a) A comparison of error thresholds and distillation rates between our distillation protocol and prior work. For comparable small codes, our protocol achieves the highest error threshold and distillation rate.
(b) The complete circuit of our  distillation protocol as described in Sec. \ref{sec:distillationprotocol} using a 2-qubit gnu code with logical states $\ket{0_{1,1,2}} = \ket{00}$ and $\ket{1_{1,1,2}} = ( \ket{10}+\ket{01})/\sqrt{2}$.
(c) Assuming a noiseless channel, the magic of the distilled state is shown as a function of the initial state parameter $v$ with $\theta = \pi/4$ for a $2,3$ and $4$-qubit code with final state given by eqs. \eqref{eq:abc2}, \eqref{eq:abc3} and \eqref{eq:abc4} respectively and magic measured by the 2-R{\'e}nyi entropy in eq. \eqref{eq:magicdensitymat}. The gray dashed and dot-dashed lines show the 2-R{\'e}nyi entropy of the states $\ket{T}$ and $\ket{H}$ respectively. Notice that, with an appropriate choice of initial state and gnu code, states with any amount of magic between $0$ and $M_2(\ket{T})$ can be distilled using our protocol, even for small codes. }
\label{Fig1}
\end{figure*}

\begin{figure*}[htbp]
\centering
\makebox[\textwidth][c]{%
    \includegraphics[width=1.05\textwidth]{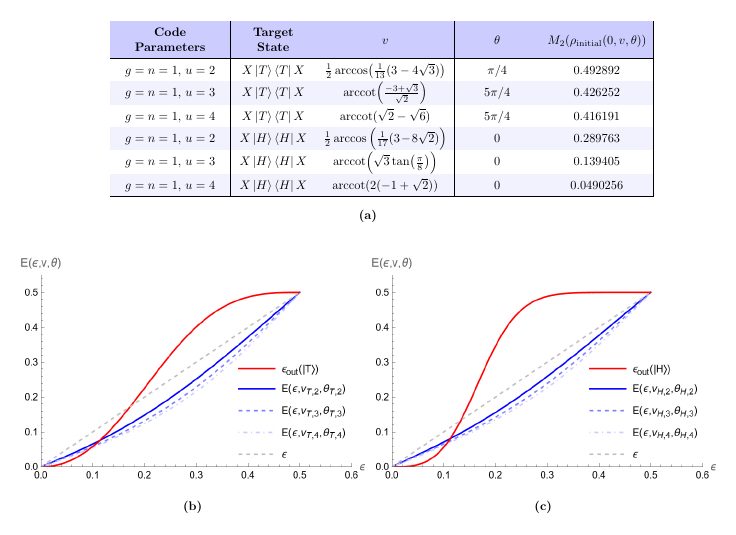}%
}
\caption{(a) Initial state parameters and their corresponding magic values required to produce the target states $X\ket{T}\bra{T}X$ or $X\ket{H}\bra{H}X$ for $2,3$ and $4$-qubit gnu codes. (b) Maximum total output error (see eq. \eqref{eq:maxerror}) for $2,3$ and $4$-qubit gnu codes (blue) for parameters presented in table (a) to achieve the final state $X\ket{T}\bra{T}X$, compared to the output error demonstrated by Bravyi and Kitaev \cite{bravyi2005universal} (red), see eq. \eqref{eq:paperTerror}. (c) Maximum total output error for $2,3$ and $4$-qubit gnu codes (blue) for parameters presented in table (a) to achieve the final state $X\ket{H}\bra{H}X$, compared to the output error achieved by Bravyi and Kitaev \cite{bravyi2005universal} (red), see eq. \eqref{eq:paperHerror}.}
\label{Fig2}
\end{figure*}

\begin{figure*}[htbp]
\centering
\makebox[\textwidth][c]{%
    \includegraphics[width=0.9\textwidth]{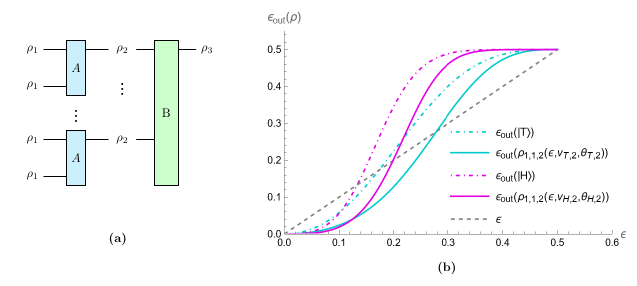}%
}
\caption{(a) We combine our protocol (labelled $A$) with existing magic state distillation protocol (labelled $B$): Protocol A produces approximate copies of the states $\ket{T}$ or $\ket{H}$ which are then used as inputs in protocol $B$ to create an even higher fidelity $\ket{T}$ or $\ket{H}$. (b) Error suppression of the combined distillation protocol in (a) for $\ket{T}$ and $\ket{H}$ (solid cyan and magenta lines, respectively) where protocol $B$ is chosen to be Bravyi and Kitaev's protocols \cite{bravyi2005universal} (see eqs. \eqref{eq:paperTerror} and \eqref{eq:paperHerror} for definitions of $\epsilon_{out}$ for $\ket{T}$ and $\ket{H}$), compared with a single iteration of protocol $B$ (dot-dashed lines). The combined protocol yields improved error suppression and higher thresholds, from approximately $0.173$ to $0.279$ for $\ket{T}$ and approximately $0.141$ to $0.198$ for $\ket{H}$.}
\label{Fig3}
\end{figure*}


Fault-tolerant quantum computation requires a universal gate set \cite{deutsch1989quantum,nielsen2010quantum}.
The most popular gate set comprises of Clifford gates, which often admit fault-tolerant implementations, and non-Clifford gates.
The non-Clifford gates are typically more challenging to implement fault-tolerantly.
Unfortunately, a universal fault-tolerant quantum computation cannot be achieved using only Clifford gates.
Magic states enable the implementation of essential non-Clifford gates via gate teleportation using Clifford gates \cite{gottesman1999demonstrating,zhou2000methodology}.
Bravyi and Kitaev \cite{bravyi2005universal} introduced the field of magic state distillation, presenting the first protocols for distilling the two most popular magic states $\ket{T}$ and $\ket{H}$.
Many noisy copies of the magic state are needed to produce a single higher-fidelity magic state, thereby suppressing errors, but at the cost of significant overhead.


Magic state distillation has been extensively studied \cite{reichardt2005quantum,reichardt2006quantum,meier2012magic,bravyi2012magic,hastings2018distillation,haah2017magic,campbell2012magic,campbell2010bound,krishna2019towards,wills2025constant,ruiz2026unfolded,hirano2025locality,zheng2025fragility,hirano2025efficient,wang2020efficiently,campbell2017unifying,cervia2025magic,erew2025pre,heussen2025magic} and many quantum error correcting codes have been considered for the construction of efficient distillation protocols, with the aim of improving both error suppression and overhead, producing higher fidelity output states from fewer noisy inputs. 
The distillation rate is the number of clean magic states produced divided by the number of noisy input states, so for some error $\epsilon$, this rate will typically be $\mathcal{O}(\log^{\gamma}(1/\epsilon))$, where $\gamma$ is the overhead exponent \cite{wills2025constant}.
Early progress showed that $\ket{H}$ could be distilled using a $4$ qubit code with both a lower $\gamma$ and higher error threshold \cite{meier2012magic} than previously achieved \cite{bravyi2005universal}, where the error threshold is the point beyond which error suppression no longer occurs.

Subsequent work demonstrated further improvements in the overhead exponent using other stabiliser codes \cite{bravyi2012magic}, punctured Reed–Muller codes \cite{hastings2018distillation}, protocols that inject additional noisy states throughout the circuit \cite{haah2017magic}, and constructions based on prime-dimensional qudits \cite{campbell2012magic,krishna2019towards}. 
Notably, using a 5-dimensional qudit Reed-Muller code \cite{campbell2012magic}, an error threshold of $\approx 0.363$ has been achieved, higher than all the aforementioned works. However, this protocol does not distil the single-qubit magic states that are the focus of this work.
Most recently, it was shown that constant rate ($\gamma = 0$) is achievable using asymptotically good triorthogonal codes \cite{wills2025constant}.
While this protocol demonstrates a good distillation rate for the state $\ket{H}$, this is only in the asymptotic limit and therefore not practical for experimental implementation. Furthermore, the error threshold tends to $0$ as the number of qubits increases.
Despite significant advancements, magic state distillation procedures are still costly both in terms of computation time and the number of qubits used \cite{campbell2017-magic-state-factory,cost-of-universality-PRXQuantum.2.020341,Litinski2019magicstate}.



Code switching \cite{poulin-code-switch-PhysRevLett.113.080501,Bombín_2016,kubica-PhysRevA.91.032330,poulsen2017fault,daguerre2025code} provides an alternative method to bypass the Eastin-Knill no-go theorem \cite{eastin2009restrictions}, which tells us that we cannot implement universal quantum gates that are fault-tolerant from being transversal on a single QEC code. While code-switching can incur significant measurement overhead \cite{cost-of-universality-PRXQuantum.2.020341}, it allows different codes to be used to implement different gates.
This can be between two stabiliser codes \cite{butt2024fault,pogorelov2025experimental,huang2023graphical,daguerre2025code}, where multiple fault-tolerant stabiliser measurements are required to transform one stabiliser code into another, which can be replaced with two single transversal measurements \cite{heussen2025efficient}. 
More recent approaches to measurement-free code-switching include using 
multi-qubit non-Clifford gates that are not transversal \cite{heussen2024measurement}, and switching between a stabiliser code that performs transversal logical Clifford gates and a permutation-invariant code that performs transversal logical non-Clifford gates \cite{ouyang2025measurement}. 
While methods to optimise code switching regimes have been demonstrated \cite{weilandt2025minimizing}, each switch is still costly as it increases the error and overhead.


The substantial qubit overhead required for both code switching and high-rate magic state distillation remains a major barrier to near-term implementation.
There is limited research on magic state distillation protocols involving fewer than the $5$ and $15$ qubit schemes originally introduced by Bravyi and Kitaev.
It has been demonstrated that magic states can be distilled by codes with as few as three qubits \cite{howard2016small,zheng2024magic} with linear error suppression, although only for a limited set of magic states such as the $|H\>$, $|T\>$ and exotic $\ket{0}+e^{i\pi/3}\ket{1}$ state.

Several experimental demonstrations of magic state distillation have now been reported, including Bravyi and Kitaev's original 5-qubit protocol being performed by an NMR quantum processor \cite{souza2011experimental}, and on a neutral atom computer \cite{sales2025experimental} after the state is encoded in 
colour codes. 
The $2$-qubit magic state $CZ$ is distilled on a superconducting qubit array using a $4$-qubit error correcting code \cite{gupta2024encoding}.
To date, magic state distillation of single-qubit magic states has not been demonstrated experimentally with fewer than five qubits. 

Other approaches to magic state distillation include multilevel distillation \cite{jones2013multilevel} and lattice surgery on 2D colour codes \cite{lee2025low}.
While most protocols allow only the distillation of specific magic states, magic dilution \cite{duclos2015reducing,campbell2016efficient} enables access to a wider range of states by reducing their magic at the cost of additional overhead on top of the distillation protocol.
In contrast, our protocol enables the direct distillation of arbitrary magic states simply by varying the input states, without requiring magic dilution.


To the best of our knowledge, all existing magic state distillation protocols are based on stabiliser codes. This motivates the question of what potential advantages might arise from distillation protocols constructed using non-stabiliser codes.


Permutation-invariant codes are generally non-stabiliser codes whose logical states are invariant under any permutation of the underlying qubits \cite{Rus00,PoR04,ouyang2014permutation,ouyang2015permutation,OUYANG201743,ouyang2019permutation,aydin2023family,ouyang2026theory}, giving them a symmetric structure is advantageous for quantum sensing applications \cite{ouyang2019robust,ouyang2022finite}, and also for the correction of both deletion \cite{ouyang2021permutation} and insertion errors \cite{bulled2025angular,bulled2026equivalence}. They form a rich family of analytically tractable codes which are also good candidates for experimental implementation in near-term devices such as trapped ions or ultracold atoms as they consist of states which are easy to prepare and can be controlled by global fields \cite{PhysRevResearch.7.L022072,PhysRevA.110.062610,johnsson2020geometric}.
As they have not yet been considered for magic state distillation, inspired by Bravyi and Kitaev \cite{bravyi2005universal}, we propose a new magic state distillation protocol that is based on permutation-invariant codes.


We focus on the gnu codes \cite{ouyang2014permutation}, a family of permutation-invariant codes whose logical states (see eq. \eqref{eq:gnustates}) alternately occupy Dicke states of higher excitation number spaced $g$ apart, with maximum occupied excitation number $gn$. The scaling parameter $u$ determines the total number of qubits $N = gnu$.
The gnu codes used in our protocol may be as small as two qubits, with the logical states $\ket{0_{1,1,2}} = \ket{00}$ and $\ket{1_{1,1,2}} = ( \ket{10}+\ket{01})/\sqrt{2}$. The 2-qubit protocol is easily performed by the simple circuit shown in Fig \ref{Fig1}(b). It requires only seven CNOT gates when we decompose the controlled-Hadamard gate with two CNOTs and some single qubit gates \cite{nielsen2010quantum}, making this a promising candidate for near-term experimental implementation.
We are the first to show a method for distilling $\ket{T}$ and $\ket{H}$ using these $2$-qubit codes, improving upon the smallest known constructions \cite{howard2016small,zheng2024magic}.
Not only does our proposed protocol have a high distillation rate of 1/2, it also achieves an error threshold of 0.5, which is higher than that reported in previous works (see Figs. \ref{Fig2}(b) and \ref{Fig2}(c) and Fig. \ref{Fig1}(a)).

Uniquely, the magic state distilled by our protocol depends only on the input states, so the protocol itself does not need to be modified, making it highly versatile. Fig. \ref{Fig1}(c) shows that states with arbitrary amounts of magic can be distilled, not just the usual $\ket{T}$ and $\ket{H}$ or the particular ``exotic'' states discussed in previous works \cite{howard2016small,zheng2024magic,duclos2013distillation}, again using permutation-invariant codes with as few as two qubits.

The improvements in error suppression and distillation rate achieved by our protocol arise from stepping outside the magic state distillation paradigm introduced by Bravyi and Kitaev \cite{bravyi2005universal}, by employing a non-Clifford circuit. Although our protocol is not fault tolerant, this is not a limitation when it is used as an initial stage prior to conventional magic state distillation. As shown in Fig.~\ref{Fig3}, combining our protocol with existing distillation schemes can greatly improve the achievable error threshold.



\section{Distilling Magic States}

Motivated by the goal of distilling, not only the commonly studied magic states $\ket{T}$ and $\ket{H}$  \cite{bravyi2005universal}, but also less explored magic states, we develop a protocol based on permutation-invariant gnu error correcting codes \cite{ouyang2014permutation}, the logical states of which are defined in \iftoggle{supp}{Supplementary Material}{Appendix} \ref{sec:distillationderivation}. These codes are particularly well suited as their simple structure allows us to determine all possible output states arising from an arbitrary input state.
In the presence of a noisy channel we define an arbitrary initial state as $\rho_{\text{initial}} = (1-\epsilon) \ket{\phi_0}\bra{\phi_0} +\epsilon\ket{\phi_1}\bra{\phi_1}$, where $\ket{\phi_0} = \cos v\ket{0} + e^{i\theta}\sin v\ket{1}$ and $\ket{\phi_1}$ is the orthogonal state to $|\phi_0\>$ (see Sec. \ref{sec:distillationprotocol}). The parameters $v$ and $\theta$ respectively describe longitude and latitude of the qubits' 
position in the Bloch sphere representation.
The parameter $\epsilon$ quantifies the error.
We then apply the projection $\Pi = \ket{0_{g,n,u}}\bra{0_{g,n,u}}+\ket{1_{g,n,u}}\bra{1_{g,n,u}}$ onto the gnu codespace, where $\ket{0_{g,n,u}}$ and $\ket{1_{g,n,u}}$ are the logical $0$ and $1$ states, to $N=gnu$ copies of the initial state. 
If our projective measurement finds the resultant state in the codespace, the protocol is successful, and otherwise, the protocol fails. 
We decode the state found in the codespace, and are left with a single-qubit final state dependent on $v$ and $\theta$ of the initial state and $g, n$ and $u$ of the corresponding code.
Sec. \ref{sec:distillationprotocol} describes these steps, with more details given in \iftoggle{supp}{Supplementary Material}{Appendix} \ref{sec:distillationderivation}. 
An analytic expression for the final state of the form eq. \eqref{eq:finalstate} for any gnu code is given by eq. \eqref{eq:generalabc}, which can be calculated efficiently with complexity $\mathcal{O}(N^3)$.

With experimental implementation in mind, we focus on the smallest gnu codes corresponding to $2,3$ and $4$ qubits, analytical expressions for the final states $\rho_{1,1,2}(\epsilon,v,\theta)$, $\rho_{1,1,3}(\epsilon,v,\theta)$ and $\rho_{1,1,4}(\epsilon,v,\theta)$ are given in \iftoggle{supp}{Supplementary Material}{Appendix} \ref{sec:appfinalstates}. 
To study the range of output states achievable, we calculate the 2-R{\'e}nyi entropy \cite{leone2022stabilizer}, which is a magic monotone \cite{veitch2014resource,seddon2021quantifying,veitch2012negative,howard2017application,wang2020efficiently,bravyi2019simulation}.
A magic monotone quantifies a state's magic and is invariant under Clifford operations (see \iftoggle{supp}{Supplementary Material}{Appendix} \ref{sec:preliminaries} for the explicit definition of $M_2(\rho)$ and $M_2(\ket{\psi})$).
For a noiseless channel and $\theta = \pi/4$, Fig. \ref{Fig1}(c) shows that varying $v$ allows us to distil a state with an arbitrary amount of magic between $0$ and $M_2(\ket{T}) \approx 0.585$.
Therefore, this protocol enables distillation of a wide range of magic states beyond $\ket{T}$ and $\ket{H}$ by varying the input state over the longitude of the Bloch sphere. Moreover, exotic magic states can be distilled using as few as two qubits via a simple circuit, shown in Fig. \ref{Fig1}(b).
The circuit takes two copies of the initial state. The first two gates, a CNOT followed by a controlled Hadamard, decodes the gnu code. The subsequent CNOT encodes the state back into the repetition code. An ancilla qubit is introduced to perform a syndrome measurement. If this fails, the state is discarded and if successful a final CNOT is applied to decode the repetition code. The second qubit is discarded, leaving the remaining qubit in the final state, $\rho_{\text{final}}$.
As summarised in Fig. \ref{Fig1}(a), this protocol represents a substantial reduction in distillation rate compared to previous approaches.

Focusing now on states $\ket{T}$ and $\ket{H}$, we aim to find the initial states required which, in the presence of noise, exhibit the largest error suppression. To investigate the output error we calculate the maximum trace distance between the final state and target state.
We are interested in the value of $\epsilon$ when it is equal to the output error of the protocol, which we define as the error threshold. Beyond the error threshold, we do not see any error suppression.
Conversely, below the error threshold, we do see error suppression.

Fig. \ref{Fig2}(a) tabulates the values of the parameters $v$ and $\theta$ that allow our protocol to distil the target states $\ket{T}$ and $\ket{H}$ using different gnu codes on 2,3 and 4 qubits. Notably, all the required initial states have lower magic than $\ket{T}$ and $\ket{H}$, respectively. 
Taking the initial states defined by Eqs. \eqref{eq:initialstates} and \eqref{eq:initial}, together with the parameters in Fig. \ref{Fig2}(a), we apply the method described in section \ref{sec:distillationprotocol} to determine the corresponding output states for each of the 2,3 and 4 qubit codes. Depending on the chosen parameters, the protocol distils either $\ket{T}$ or $\ket{H}$. 
Figures \ref{Fig2}(b) and \ref{Fig2}(c) show the maximum output error (see Eq. \ref{eq:maxerror}) as a function of the input error $\epsilon$ for the $2,3$ and $4$-qubit gnu codes, for $\ket{T}$ and $\ket{H}$, respectively.
We compare our output errors with the output error achieved using the 5-qubit and 15-qubit distillation protocols \cite{bravyi2005universal} (in red). 
While Bravyi and Kitaev's protocols \cite{bravyi2005universal} result in a lower output error for very small $\epsilon$, for $\epsilon>0.114$ or $\epsilon>0.112$ our 2-qubit code shows better error suppression for $\ket{T}$ and $\ket{H}$ respectively while requiring fewer physical qubits. 
Also, our error threshold is $0.5$ which is much larger than $0.173$ or $0.141$ for the states $\ket{T}$ and $\ket{H}$ respectively \cite{bravyi2005universal}, or $0.146$ for $\ket{H}$ achieved by Reichardt \cite{reichardt2006quantum}. Figs. \ref{Fig2}(b) and \ref{Fig2}(c) show this substantial improvement in error threshold compared to the other similar protocols summarised in Fig. \ref{Fig2}(a).

\section{Incorporating Existing Magic State Distillation Protocols}
One of the main benefits of our protocol is that it can be used alongside existing magic state distillation protocols as in Fig. \ref{Fig3}(a). Focusing on our 2-qubit protocol, shown in the circuit Fig. \ref{Fig1}(b), as it has the best distillation rate amongst the gnu codes we explored. We will label this protocol $A$.
For convenience, we label Bravyi and Kitaev's original magic state distillation protocols as $B$.
If protocol $A$ is used first to distil the $5$ or $15$ noisy states that are required for protocol $B$ to distil a $\ket{T}$ or $\ket{H}$, then we only need $2\times 5$ or $2\times 15$ qubits in total, whereas two iterations of protocol $B$ would take $5^2$ or $15^2$ qubits. 
The combined protocol $A+B$ only requires doubling the number of qubits in protocol $B$, but as seen in Fig. \ref{Fig3}(b), this greatly increases the error threshold, from approximately $0.173$ to $0.279$ for $\ket{T}$ and approximately $0.141$ to $0.198$ for $\ket{H}$. 
In Fig \ref{Fig3}(b) we plot the output error of our 2-qubit protocol as a function of the input error $\epsilon$, for the distillation of the magic states $\ket{T}$ and $\ket{H}$. The final states are given by Eq. \eqref{eq:abc2}, with parameters specified in Fig. \ref{Fig2}(a). 
These output states are then used as inputs to the Bravyi–Kitaev protocols, yielding the total output errors shown by the solid cyan ($\ket{T}$) and magenta ($\ket{H}$) lines. 
For comparison, the cyan and magenta dot-dashed lines show the output error (Eqs. \eqref{eq:paperTerror} and \eqref{eq:paperHerror}) obtained from a single iteration of the Bravyi–Kitaev protocol for distilling $\ket{T}$ and $\ket{H}$, respectively.
While this example uses the Bravyi and Kitaev protocol \cite{bravyi2005universal}, protocol $A$ can easily be incorporated with any magic state distillation protocol to produce higher fidelity initial states for protocol $B$, improving the error suppression, 
while only doubling the number of qubits required.

\section{Magic State Distillation Protocol}\label{sec:distillationprotocol}
Consider orthonormal states 
\begin{equation}
\begin{split}\label{eq:initialstates}
\ket{\phi_0} &= \cos v\ket{0} + e^{i\theta}\sin v\ket{1}, \\
\ket{\phi_1} &= \sin v\ket{0} - e^{i\theta}\cos v\ket{1},
\end{split}
\end{equation}
so that in the presence of a noisy channel we have an initial state that takes the form of the density matrix
\begin{equation}\label{eq:initial}
\rho_{\text{initial},\epsilon} = (1-\epsilon) \ket{\phi_0}\bra{\phi_0} +\epsilon\ket{\phi_1}\bra{\phi_1}.
\end{equation}
We then take $N$ identical copies of the initial state
\begin{align}
\rho_{N,\epsilon}=
\sum_{{\bf x} \in \{0,1\}^N}
    \epsilon^{|{\bf x}|}(1-\epsilon )^{N-|{\bf x}|} |\phi_{\bf x}\>\<\phi_{\bf x}|,
\end{align}
where, for a binary string $\textbf{x} \in \{0,1\}^N$, we write the tensor product state $ \ket{\phi_{\bf x}} =\ket{\phi_{x_1}} \otimes \dots \otimes \ket{\phi_{x_N}}$. We will then apply the projection 
\begin{equation}
\Pi = \ket{0_{g,n,u}}\bra{0_{g,n,u}}+\ket{1_{g,n,u}}\bra{1_{g,n,u}}
\end{equation}
so that the resulting $N$-qubit state, after normalising, is of the form
\begin{equation}\label{eq:finalstate}
\rho_{\text{final},\epsilon} =\frac{\Pi \rho_{N,\epsilon} \Pi}{\Tr[\Pi \rho_{N,\epsilon} \Pi]}  = \frac{1}{a+b}\begin{pmatrix} a & c\\
c^* & b
\end{pmatrix} .
\end{equation}
where $a,b$ and $c$ are all functions of $v, \theta$ and $\epsilon$. 
We can simplify $a,b$ and $c$ as $\langle D_w^N | \phi_{\bf x} \rangle$ admits the analytical expression 
\begin{equation}
\langle D_s^N|\phi_{\textbf{x}}\rangle = \frac{1}{\sqrt{\binom{N}{s}}} \sum_{t=\max(0,s+|{\bf x}|-N)}^{\min(s,|\textbf{x}|)} \alpha_{s,t} \binom{|\textbf{x}|}{t}\binom{N-|\textbf{x}|}{s-t}.\label{eq:Dickeinnerproduct}
\end{equation}
where $|\textbf{x}|$ denotes the Hamming weight of ${\bf x}$, $N=gnu$ and 
\begin{equation}
\begin{split}
\alpha_{s,t} 
& = (-1)^te^{is\theta}(\cos v)^N(\tan v)^{s+|\textbf{x}|-2t}.
\end{split}
\end{equation}
A detailed derivation of $a,b$ and $c$ is given in \iftoggle{supp}{Supplementary Material}{Appendix} \ref{sec:distillationderivation}, along with their explicit values for the $2,3$ and $4$ qubit codes in \iftoggle{supp}{Supplementary Material}{Appendix} \ref{sec:appfinalstates}.
The complexity of evaluating the expression in \eqref{eq:Dickeinnerproduct} is $O(N^2)$, and therefore, the overall complexity of evaluating the values of $a$, $b$ and $c$ is $O(N^3)$. 

We quantify the proximity of the distilled state to the ideal target state $\rho_{\text{target}}$ by computing the trace distance $D(\rho_{\text{final},\epsilon},\rho_{\text{target}}) = \frac{1}{2}\|\rho_{\text{final},\epsilon} - \rho_{\text{target}}\|_1$, so that the corresponding maximum error is is given by
\begin{equation}\label{eq:maxerror}
    E = \max_{e \in \{0,\epsilon\}} D(\rho_{\text{final},e}, \rho_{\text{target}}  ).
\end{equation}
The probability of success of the distillation protocol is just $\tr[\Pi \rho_{N,\epsilon} \Pi]$ which is 
\begin{equation}\label{eq:ps}
   p_s = a+b.
\end{equation}

\section{Discussions}\label{sec:conclusion}
We have presented an alternative magic state distillation protocol that uses permutation-invariant codes. We were able to show examples that require as few as two qubits (the gnu code with $g = n = 1$ and $u=2$) which is somewhat surprising, since codes of this size are typically considered too small to be useful for error correction.
Despite this, they can still be used to distil both $\ket{T}$ and $\ket{H}$ with a strong error suppression, even beyond the threshold error shown by Bravyi and Kitaev \cite{bravyi2005universal} and Reichardt \cite{reichardt2006quantum} while using fewer qubits. These small codes can also be used to distil a wide range of exotic magic states, beyond the examples shown previously \cite{howard2016small,zheng2024magic,duclos2013distillation}.

Although our protocol is not fault-tolerant, it can be used as a base-layer with existing fault-tolerant magic state distillation protocols, as in Figs. \ref{Fig3}(a) and \ref{Fig3}(b) to greatly improve their error suppression with only a linear increase in distillation rate. 
The extremely low number of gates that our protocol requires (see Fig.\ref{Fig1}(b)) is also particularly advantageous for its near-term implementation, even with imperfect gates. 
Several open questions remain for future investigation.
While we have restricted our analysis here to gnu codes, we expect that similar behaviour may arise for other families of permutation-invariant codes, we would like to investigate whether these other families of codes can provide similar or possibly enhanced results. 

An immediate question that comes to mind is whether we can use stabiliser codes to also obtain a similar result. Bravyi and Kitaev \cite[Section III]{bravyi2005universal} proposed a magic state dilution protocol that uses a two-qubit repetition code to produce a $(|0\> + e^{-i\pi/6}|1\>)/\sqrt 2$ magic state from two copies of $T$-states. 
Since the two-qubit repetition code is gnu code, we can use our formalism to study its performance.
We find that if this two-qubit repetition code were used to distil the $T$ and $H$ magic states, the output error is unfortunately larger than the input error for all positive errors (see \iftoggle{supp}{Supplementary Material}{Appendix} \ref{sec:2rep}). 

Additionally, developing a method of quantifying the advantage of these permutation-invariant code based protocols would be valuable so that we could compare their performance both alone and in combination with other magic state distillation protocols.


\section{Acknowledgements}\label{eq:acknow}
Y.O. and H.L. acknowledge support from EPSRC (Grant No. EP/W028115/1). 
Y.O. also acknowledges support from the EPSRC funded QCI3 Hub under Grant No. EP/Z53318X/1.  

\bibliography{bib}

@article{zheng2024magic,
  doi = {10.22331/q-2025-09-15-1858},
  url = {https://doi.org/10.22331/q-2025-09-15-1858},
  title = {From {M}agic {S}tate {D}istillation to {D}ynamical {S}ystems},
  author = {Zheng, Yunzhe and Liu, Dong E.},
  journal = {{Quantum}},
  issn = {2521-327X},
  publisher = {{Verein zur F{\"{o}}rderung des Open Access Publizierens in den Quantenwissenschaften}},
  volume = {9},
  pages = {1858},
  month = sep,
  year = {2025}
}

@article{leone2022stabilizer,
  title={Stabilizer r{\'e}nyi entropy},
  author={Leone, Lorenzo and Oliviero, Salvatore FE and Hamma, Alioscia},
  journal={Physical Review Letters},
  volume={128},
  number={5},
  pages={050402},
  year={2022},
  publisher={APS}
}

@article{howard2016small,
  title={Small codes for magic state distillation},
  author={Howard, Mark and Dawkins, Hillary},
  journal={The European Physical Journal D},
  volume={70},
  number={3},
  pages={55},
  year={2016},
  publisher={Springer}
}

@article{ouyang2014permutation,
  title={Permutation-invariant quantum codes},
  author={Ouyang, Yingkai},
  journal={Physical Review A},
  volume={90},
  number={6},
  pages={062317},
  year={2014},
  publisher={APS}
}

@inproceedings{ouyang2021permutation,
  title={Permutation-invariant quantum coding for quantum deletion channels},
  author={Ouyang, Yingkai},
  booktitle={2021 IEEE International Symposium on Information Theory (ISIT)},
  pages={1499--1503},
  year={2021},
  organization={IEEE}
}

@article{bravyi2005universal,
  title={Universal quantum computation with ideal Clifford gates and noisy ancillas},
  author={Bravyi, Sergey and Kitaev, Alexei},
  journal={Physical Review A—Atomic, Molecular, and Optical Physics},
  volume={71},
  number={2},
  pages={022316},
  year={2005},
  publisher={APS}
}

@article{meier2012magic,
  title={Magic-state distillation with the four-qubit code},
  author={Meier, Adam M and Eastin, Bryan and Knill, Emanuel},
  journal={arXiv preprint arXiv:1204.4221},
  year={2012}
}

@article{wills2025constant,
  title={Constant-overhead magic state distillation},
  author={Wills, Adam and Hsieh, Min-Hsiu and Yamasaki, Hayata},
  journal={Nature Physics},
  volume={21},
  number={11},
  pages={1842--1846},
  year={2025},
  publisher={Nature Publishing Group UK London}
}

@article{bravyi2012magic,
  title={Magic-state distillation with low overhead},
  author={Bravyi, Sergey and Haah, Jeongwan},
  journal={Physical Review A—Atomic, Molecular, and Optical Physics},
  volume={86},
  number={5},
  pages={052329},
  year={2012},
  publisher={APS}
}

@article{hastings2018distillation,
  title={Distillation with sublogarithmic overhead},
  author={Hastings, Matthew B and Haah, Jeongwan},
  journal={Physical review letters},
  volume={120},
  number={5},
  pages={050504},
  year={2018},
  publisher={APS}
}

@article{sales2025experimental,
  title={Experimental demonstration of logical magic state distillation},
  author={Sales Rodriguez, Pedro and Robinson, John M and Jepsen, Paul Niklas and He, Zhiyang and Duckering, Casey and Zhao, Chen and Wu, Kai-Hsin and Campo, Joseph and Bagnall, Kevin and Kwon, Minho and others},
  journal={Nature},
  volume={645},
  number={8081},
  pages={620--625},
  year={2025},
  publisher={Nature Publishing Group UK London}
}

@article{haah2017magic,
  title={Magic state distillation with low space overhead and optimal asymptotic input count},
  author={Haah, Jeongwan and Hastings, Matthew B and Poulin, David and Wecker, D},
  journal={Quantum},
  volume={1},
  pages={31},
  year={2017},
  publisher={Verein zur F{\"o}rderung des Open Access Publizierens in den Quantenwissenschaften}
}

@article{jones2013multilevel,
  title={Multilevel distillation of magic states for quantum computing},
  author={Jones, Cody},
  journal={Physical Review A—Atomic, Molecular, and Optical Physics},
  volume={87},
  number={4},
  pages={042305},
  year={2013},
  publisher={APS}
}

@article{lee2025low,
  title={Low-overhead magic state distillation with color codes},
  author={Lee, Seok-Hyung and Thomsen, Felix and Fazio, Nicholas and Brown, Benjamin J and Bartlett, Stephen D},
  journal={PRX Quantum},
  volume={6},
  number={3},
  pages={030317},
  year={2025},
  publisher={APS}
}

@article{souza2011experimental,
  title={Experimental magic state distillation for fault-tolerant quantum computing},
  author={Souza, Alexandre M and Zhang, Jingfu and Ryan, Colm A and Laflamme, Raymond},
  journal={Nature communications},
  volume={2},
  number={1},
  pages={169},
  year={2011},
  publisher={Nature Publishing Group UK London}
}

@article{gupta2024encoding,
  title={Encoding a magic state with beyond break-even fidelity},
  author={Gupta, Riddhi S and Sundaresan, Neereja and Alexander, Thomas and Wood, Christopher J and Merkel, Seth T and Healy, Michael B and Hillenbrand, Marius and Jochym-O’Connor, Tomas and Wootton, James R and Yoder, Theodore J and others},
  journal={Nature},
  volume={625},
  number={7994},
  pages={259--263},
  year={2024},
  publisher={Nature Publishing Group UK London}
}

@article{daguerre2025code,
  title={Code switching revisited: Low-overhead magic state preparation using color codes},
  author={Daguerre, Lucas and Kim, Isaac H},
  journal={Physical Review Research},
  volume={7},
  number={2},
  pages={023080},
  year={2025},
  publisher={APS}
}

@article{reichardt2005quantum,
  title={Quantum universality from magic states distillation applied to CSS codes},
  author={Reichardt, Ben W},
  journal={Quantum Information Processing},
  volume={4},
  number={3},
  pages={251--264},
  year={2005},
  publisher={Springer}
}

@article{reichardt2006quantum,
author = {Reichardt, Ben W.},
title = {Quantum universality by state distillation},
year = {2009},
issue_date = {November 2009},
publisher = {Rinton Press, Incorporated},
address = {Paramus, NJ},
volume = {9},
number = {11},
issn = {1533-7146},
journal = {Quantum Info. Comput.},
month = nov,
pages = {1030–1052},
numpages = {23}
}

@article{duclos2013distillation,
  title={Distillation of nonstabilizer states for universal quantum computation},
  author={Duclos-Cianci, Guillaume and Svore, Krysta M},
  journal={Physical Review A—Atomic, Molecular, and Optical Physics},
  volume={88},
  number={4},
  pages={042325},
  year={2013},
  publisher={APS}
}

@article{eastin2009restrictions,
  title={Restrictions on transversal encoded quantum gate sets},
  author={Eastin, Bryan and Knill, Emanuel},
  journal={Physical review letters},
  volume={102},
  number={11},
  pages={110502},
  year={2009},
  publisher={APS}
}

@article{ouyang2025measurement,
  title={Measurement-free code-switching protocol for low-overhead quantum computation using permutation-invariant codes},
  author={Ouyang, Yingkai and Jing, Yumang and Brennen, Gavin K},
  journal={PRX Quantum},
  volume={6},
  number={4},
  pages={040341},
  year={2025},
  publisher={APS}
}

@article{PhysRevResearch.7.L022072,
  title = {Global variational quantum circuits for arbitrary symmetric state preparation},
  author = {Bond, Liam J. and Davis, Matthew J. and Min\'a\ifmmode \check{r}\else \v{r}\fi{}, Ji\ifmmode \check{r}\else \v{r}\fi{}\'{\i} and Gerritsma, Rene and Brennen, Gavin K. and Safavi-Naini, Arghavan},
  journal = {Phys. Rev. Res.},
  volume = {7},
  issue = {2},
  pages = {L022072},
  numpages = {7},
  year = {2025},
  month = {Jun},
  publisher = {American Physical Society},
  doi = {10.1103/PhysRevResearch.7.L022072},
  url = {https://link.aps.org/doi/10.1103/PhysRevResearch.7.L022072}
}

@article{heussen2024measurement,
  title={Measurement-free fault-tolerant quantum error correction in near-term devices},
  author={Heu{\ss}en, Sascha and Locher, David F and M{\"u}ller, Markus},
  journal={PRX Quantum},
  volume={5},
  number={1},
  pages={010333},
  year={2024},
  publisher={APS},
doi={10.1103/PRXQuantum.5.010333}
}

@article{campbell2017-magic-state-factory,
  title = {Quantum computation with realistic magic-state factories},
  author = {O'Gorman, Joe and Campbell, Earl T.},
  journal = {Phys. Rev. A},
  volume = {95},
  issue = {3},
  pages = {032338},
  numpages = {19},
  year = {2017},
  month = {Mar},
  publisher = {American Physical Society},
  doi = {10.1103/PhysRevA.95.032338},
  url = {https://link.aps.org/doi/10.1103/PhysRevA.95.032338}
}

@article{poulsen2017fault,
  title={Fault-tolerant interface between quantum memories and quantum processors},
  author={Poulsen Nautrup, Hendrik and Friis, Nicolai and Briegel, Hans J},
  journal={Nature communications},
  volume={8},
  number={1},
  pages={1321},
  year={2017},
  publisher={Nature Publishing Group UK London},
  doi={10.1038/s41467-017-01418-2}
}

@article{cost-of-universality-PRXQuantum.2.020341,
  title = {Cost of Universality: A Comparative Study of the Overhead of State Distillation and Code Switching with Color Codes},
  author = {Beverland, Michael E. and Kubica, Aleksander and Svore, Krysta M.},
  journal = {PRX Quantum},
  volume = {2},
  issue = {2},
  pages = {020341},
  numpages = {46},
  year = {2021},
  month = {Jun},
  publisher = {American Physical Society},
  doi = {10.1103/PRXQuantum.2.020341},
  url = {https://link.aps.org/doi/10.1103/PRXQuantum.2.020341}
}

@article{krishna2019towards,
  title = {Towards Low Overhead Magic State Distillation},
  author = {Krishna, Anirudh and Tillich, Jean-Pierre},
  journal = {Phys. Rev. Lett.},
  volume = {123},
  issue = {7},
  pages = {070507},
  numpages = {4},
  year = {2019},
  month = {Aug},
  publisher = {American Physical Society},
  doi = {10.1103/PhysRevLett.123.070507},
  url = {https://link.aps.org/doi/10.1103/PhysRevLett.123.070507}
}

@article{campbell2017unifying,
  title = {Unifying Gate Synthesis and Magic State Distillation},
  author = {Campbell, Earl T. and Howard, Mark},
  journal = {Phys. Rev. Lett.},
  volume = {118},
  issue = {6},
  pages = {060501},
  numpages = {5},
  year = {2017},
  month = {Feb},
  publisher = {American Physical Society},
  doi = {10.1103/PhysRevLett.118.060501},
  url = {https://link.aps.org/doi/10.1103/PhysRevLett.118.060501}
}

@article{Litinski2019magicstate,
  doi = {10.22331/q-2019-12-02-205},
  url = {https://doi.org/10.22331/q-2019-12-02-205},
  title = {Magic {S}tate {D}istillation: {N}ot as {C}ostly as {Y}ou {T}hink},
  author = {Litinski, Daniel},
  journal = {{Quantum}},
  issn = {2521-327X},
  publisher = {{Verein zur F{\"{o}}rderung des Open Access Publizierens in den Quantenwissenschaften}},
  volume = {3},
  pages = {205},
  month = {dec},
  year = {2019}
}

@article{poulin-code-switch-PhysRevLett.113.080501,
  title = {Fault-Tolerant Conversion between the Steane and Reed-Muller Quantum Codes},
  author = {Anderson, Jonas T. and Duclos-Cianci, Guillaume and Poulin, David},
  journal = {Phys. Rev. Lett.},
  volume = {113},
  issue = {8},
  pages = {080501},
  numpages = {5},
  year = {2014},
  month = {Aug},
  publisher = {American Physical Society},
  doi = {10.1103/PhysRevLett.113.080501},
  url = {https://link.aps.org/doi/10.1103/PhysRevLett.113.080501}
}

@article{Bombín_2016,
doi = {10.1088/1367-2630/18/4/043038},
url = {https://dx.doi.org/10.1088/1367-2630/18/4/043038},
year = {2016},
month = {apr},
publisher = {IOP Publishing},
volume = {18},
number = {4},
pages = {043038},
author = {Héctor Bombín},
title = {Dimensional jump in quantum error correction},
journal = {New Journal of Physics}
}

@article{kubica-PhysRevA.91.032330,
  title = {Universal transversal gates with color codes: A simplified approach},
  author = {Kubica, Aleksander and Beverland, Michael E.},
  journal = {Phys. Rev. A},
  volume = {91},
  issue = {3},
  pages = {032330},
  numpages = {12},
  year = {2015},
  month = {Mar},
  publisher = {American Physical Society},
  doi = {10.1103/PhysRevA.91.032330},
  url = {https://link.aps.org/doi/10.1103/PhysRevA.91.032330}
}

@article{ouyang2019robust,
  author={Ouyang, Yingkai and Shettell, Nathan and Markham, Damian},
  journal={IEEE Transactions on Information Theory}, 
  title={Robust Quantum Metrology With Explicit Symmetric States}, 
  year={2022},
  volume={68},
  number={3},
  pages={1809-1821},
  doi={10.1109/TIT.2021.3132634}}

@article{heussen2025efficient,
  title={Efficient fault-tolerant code switching via one-way transversal CNOT gates},
  author={Heu{\ss}en, Sascha and Hilder, Janine},
  journal={Quantum},
  volume={9},
  pages={1846},
  year={2025},
  publisher={Verein zur F{\"o}rderung des Open Access Publizierens in den Quantenwissenschaften}
}

@article{butt2024fault,
  title={Fault-Tolerant Code-Switching Protocols for Near-Term Quantum Processors},
  author={Butt, Friederike and Heu{\ss}en, Sascha and Rispler, Manuel and M{\"u}ller, Markus},
  journal={PRX Quantum},
  volume={5},
  number={2},
  pages={020345},
  year={2024},
  publisher={APS},
  doi={10.1103/PRXQuantum.5.020345}
}

@article{pogorelov2025experimental,
  title={Experimental fault-tolerant code switching},
  author={Pogorelov, Ivan and Butt, Friederike and Postler, Lukas and Marciniak, Christian D and Schindler, Philipp and M{\"u}ller, Markus and Monz, Thomas},
  journal={Nature Physics},
  volume={21},
  number={2},
  pages={298--303},
  year={2025},
  publisher={Nature Publishing Group UK London}
}

@article{huang2023graphical,
  title={Graphical {CSS} code transformation using {ZX} calculus},
  author={Huang, Jiaxin and Li, Sarah Meng and Yeh, Lia and Kissinger, Aleks and Mosca, Michele and Vasmer, Michael},
  journal={arXiv preprint arXiv:2307.02437},
  year={2023},
  url={https://arxiv.org/abs/2307.02437}
}

@article{campbell2012magic,
  title={Magic-state distillation in all prime dimensions using quantum reed-muller codes},
  author={Campbell, Earl T and Anwar, Hussain and Browne, Dan E},
  journal={Physical Review X},
  volume={2},
  number={4},
  pages={041021},
  year={2012},
  publisher={APS}
}

@article{weilandt2025minimizing,
  title={Minimizing the Number of Code Switching Operations in Fault-Tolerant Quantum Circuits},
  author={Weilandt, Erik and Peham, Tom and Wille, Robert},
  journal={arXiv preprint arXiv:2512.04170},
  year={2025}
}

@article{bulled2026equivalence,
  title={The equivalence of quantum deletion and insertion errors on permutation-invariant codes},
  author={Bulled, Lewis and Ouyang, Yingkai},
  journal={arXiv preprint arXiv:2602.08780},
  year={2026}
}

@article{gottesman1999demonstrating,
  title={Demonstrating the viability of universal quantum computation using teleportation and single-qubit operations},
  author={Gottesman, Daniel and Chuang, Isaac L},
  journal={Nature},
  volume={402},
  number={6760},
  pages={390--393},
  year={1999},
  publisher={Nature Publishing Group UK London}
}

@article{zhou2000methodology,
  title={Methodology for quantum logic gate construction},
  author={Zhou, Xinlan and Leung, Debbie W and Chuang, Isaac L},
  journal={Physical Review A},
  volume={62},
  number={5},
  pages={052316},
  year={2000},
  publisher={APS}
}

@article{campbell2010bound,
  title={Bound states for magic state distillation in fault-tolerant quantum computation},
  author={Campbell, Earl T and Browne, Dan E},
  journal={Physical review letters},
  volume={104},
  number={3},
  pages={030503},
  year={2010},
  publisher={APS}
}

@article{deutsch1989quantum,
  title={Quantum computational networks},
  author={Deutsch, David Elieser},
  journal={Proceedings of the royal society of London. A. mathematical and physical sciences},
  volume={425},
  number={1868},
  pages={73--90},
  year={1989},
  publisher={The Royal Society London}
}

@book{nielsen2010quantum,
  title={Quantum computation and quantum information},
  author={Nielsen, Michael A and Chuang, Isaac L},
  year={2010},
  publisher={Cambridge university press}
}

@article{ruiz2026unfolded,
  title={Unfolded distillation: very low-cost magic state preparation for biased-noise qubits},
  author={Ruiz, Diego and Guillaud, J{\'e}r{\'e}mie and Vuillot, Christophe and Mirrahimi, Mazyar},
  journal={npj Quantum Information},
  year={2026},
  publisher={Nature Publishing Group UK London}
}

@inproceedings{hirano2025locality,
  title={Locality-aware Pauli-based computation for local magic state preparation},
  author={Hirano, Yutaka and Fujii, Keisuke},
  booktitle={2025 IEEE International Conference on Quantum Computing and Engineering (QCE)},
  volume={1},
  pages={670--680},
  year={2025},
  organization={IEEE}
}

@article{zheng2025fragility,
  title={Fragility of Magic State Distillation under Imperfect Measurements},
  author={Zheng, Yunzhe and Zhao, Yuanchen and Liu, Dong E},
  journal={arXiv preprint arXiv:2503.01165},
  year={2025}
}

@article{hirano2025efficient,
  title={Efficient magic state cultivation with lattice surgery},
  author={Hirano, Yutaka and Toshio, Riki and Itogawa, Tomohiro and Fujii, Keisuke},
  journal={arXiv preprint arXiv:2510.24615},
  year={2025}
}

@article{wang2020efficiently,
  title={Efficiently computable bounds for magic state distillation},
  author={Wang, Xin and Wilde, Mark M and Su, Yuan},
  journal={Physical review letters},
  volume={124},
  number={9},
  pages={090505},
  year={2020},
  publisher={APS}
}

@article{campbell2016efficient,
  title={An efficient magic state approach to small angle rotations},
  author={Campbell, Earl T and O’Gorman, Joe},
  journal={Quantum Science and Technology},
  volume={1},
  number={1},
  pages={015007},
  year={2016},
  publisher={IOP Publishing}
}

@article{duclos2015reducing,
  title={Reducing the quantum-computing overhead with complex gate distillation},
  author={Duclos-Cianci, Guillaume and Poulin, David},
  journal={Physical Review A},
  volume={91},
  number={4},
  pages={042315},
  year={2015},
  publisher={APS}
}

@article{seddon2021quantifying,
  title={Quantifying quantum speedups: Improved classical simulation from tighter magic monotones},
  author={Seddon, James R and Regula, Bartosz and Pashayan, Hakop and Ouyang, Yingkai and Campbell, Earl T},
  journal={PRX Quantum},
  volume={2},
  number={1},
  pages={010345},
  year={2021},
  publisher={APS}
}

@article{johnsson2020geometric,
  title={Geometric Pathway to Scalable Quantum Sensing},
  author={Johnsson, Mattias T and Mukty, Nabomita Roy and Burgarth, Daniel and Volz, Thomas and Brennen, Gavin K},
  journal={Physical Review Letters},
  volume={125},
  number={19},
  pages={190403},
  year={2020},
  publisher={APS},
doi={10.1103/PhysRevLett.125.190403}
}

@article{PhysRevA.110.062610,
  title = {Nonlocal multiqubit quantum gates via a driven cavity},
  author = {Jandura, Sven and Srivastava, Vineesha and Pecorari, Laura and Brennen, Gavin K. and Pupillo, Guido},
  journal = {Phys. Rev. A},
  volume = {110},
  issue = {6},
  pages = {062610},
  numpages = {17},
  year = {2024},
  month = {Dec},
  publisher = {American Physical Society},
  doi = {10.1103/PhysRevA.110.062610},
  url = {https://link.aps.org/doi/10.1103/PhysRevA.110.062610}
}

@article{PoR04,
author = {Pollatsek, Harriet and Ruskai, Mary Beth},
doi = {10.1016/j.laa.2004.06.014},
issn = {0024-3795},
journal = {Linear Algebra and its Applications},
number = {0},
pages = {255--288},
title = {{Permutationally invariant codes for quantum error correction}},
url = {http://www.sciencedirect.com/science/article/pii/S0024379504002903},
volume = {392},
year = {2004}
}

@article{Rus00,
author = {Ruskai, Mary Beth},
doi = {10.1103/PhysRevLett.85.194},
journal = {Physical Review Letters},
month = jul,
number = {1},
pages = {194--197},
publisher = {American Physical Society},
title = {{Pauli Exchange Errors in Quantum Computation}},
url = {http://link.aps.org/doi/10.1103/PhysRevLett.85.194},
volume = {85},
year = {2000}
}

@article{ouyang2019permutation,
  title={Permutation-invariant constant-excitation quantum codes for amplitude damping},
  author={Ouyang, Yingkai and Chao, Rui},
  doi={10.1109/TIT.2019.2956142},
  journal={IEEE Transactions on Information Theory},
  volume={66},
  number={5},
  pages={2921--2933},
  year={2019},
  publisher={IEEE}
}

@article{OUYANG201743,
title = "Permutation-invariant qudit codes from polynomials",
journal = "Linear Algebra and its Applications",
volume = "532",
number = "",
pages = "43 - 59",
year = "2017",
note = "",
issn = "0024-3795",
doi = "http://dx.doi.org/10.1016/j.laa.2017.06.031",
url = "http://www.sciencedirect.com/science/article/pii/S0024379517303956",
author = "Yingkai Ouyang"
}

@article{ouyang2015permutation,
  title = {Permutation-invariant codes encoding more than one qubit},
  author = {Ouyang, Yingkai and Fitzsimons, Joseph},
  journal = {Physical Review A},
  volume = {93},
  issue = {4},
  pages = {042340},
  numpages = {4},
  year = {2016},
  month = {Apr},
  publisher = {American Physical Society},
  doi = {10.1103/PhysRevA.93.042340},
  url = {http://link.aps.org/doi/10.1103/PhysRevA.93.042340}
}

@article{aydin2023family,  doi = {10.22331/q-2024-04-30-1321},
  url = {https://doi.org/10.22331/q-2024-04-30-1321},
  title = {A family of permutationally invariant quantum codes},
  author = {Aydin, Arda and Alekseyev, Max A. and Barg, Alexander},
  journal = {{Quantum}},
  issn = {2521-327X},
  publisher = {{Verein zur F{\"{o}}rderung des Open Access Publizierens in den Quantenwissenschaften}},
  volume = {8},
  pages = {1321},
  month = {apr},
  year = {2024}
}

@article{ouyang2026theory,
  title={A theory of quantum error correction for permutation-invariant codes},
  author={Ouyang, Yingkai and Brennen, Gavin K},
  journal={arXiv preprint arXiv:2602.13638},
  year={2026}
}

@article{ouyang2022finite,
  title={Finite-round quantum error correction on symmetric quantum sensors},
  author={Ouyang, Yingkai and Brennen, Gavin K},
  journal={arXiv preprint arXiv:2212.06285},
  year={2022}
}

@article{bulled2025angular,
  title={An angular momentum approach to quantum insertion errors},
  author={Bulled, Lewis and Ouyang, Yingkai},
  journal={arXiv preprint arXiv:2509.03413},
  year={2025}
}

@article{cervia2025magic,
  title={Magic State Distillation using Asymptotically Good Codes on Qudits},
  author={Cervia, Michael J and Lamm, Henry and Liu, Diyi and Murairi, Edison M and Zhu, Shuchen},
  journal={arXiv preprint arXiv:2512.21874},
  year={2025}
}

@article{erew2025pre,
  title={Pre-Distillation of Magic States via Composite Schemes},
  author={Erew, Muhammad and Goldstein, Moshe and Oz, Yaron and Suchowski, Haim},
  journal={arXiv preprint arXiv:2510.00804},
  year={2025}
}

@article{heussen2025magic,
  title={Magic state distillation without measurements and post-selection},
  author={Heu{\ss}en, Sascha},
  journal={APL Quantum},
  volume={2},
  number={4},
  year={2025},
  publisher={AIP Publishing}
}

@article{veitch2014resource,
  title={The resource theory of stabilizer quantum computation},
  author={Veitch, Victor and Hamed Mousavian, SA and Gottesman, Daniel and Emerson, Joseph},
  journal={New Journal of Physics},
  volume={16},
  number={1},
  pages={013009},
  year={2014},
  publisher={IOP Publishing}
}

@article{veitch2012negative,
  title={Negative quasi-probability as a resource for quantum computation},
  author={Veitch, Victor and Ferrie, Christopher and Gross, David and Emerson, Joseph},
  journal={New Journal of Physics},
  volume={14},
  number={11},
  pages={113011},
  year={2012},
  publisher={IOP Publishing}
}

@article{howard2017application,
  title={Application of a resource theory for magic states to fault-tolerant quantum computing},
  author={Howard, Mark and Campbell, Earl},
  journal={Physical review letters},
  volume={118},
  number={9},
  pages={090501},
  year={2017},
  publisher={APS}
}

@article{bravyi2019simulation,
  title={Simulation of quantum circuits by low-rank stabilizer decompositions},
  author={Bravyi, Sergey and Browne, Dan and Calpin, Padraic and Campbell, Earl and Gosset, David and Howard, Mark},
  journal={Quantum},
  volume={3},
  pages={181},
  year={2019},
  publisher={Verein zur F{\"o}rderung des Open Access Publizierens in den Quantenwissenschaften}
}

\iftoggle{supp}{\appendix\section*{Supplementary Material}\renewcommand{\thesection}{S \Alph{section}}\renewcommand{\appendixname}{}}{\appendix}

\section{Preliminaries}\label{sec:preliminaries}
\subsection{Magic states}
We will focus on the two most common magic states \cite{bravyi2005universal}, $T$-type
\begin{equation}
     \ket{T} = \cos\beta \ket{0} + e^{i\pi/4}\sin\beta\ket{1} , \;\;\;\; \; \cos(2\beta) = \frac{1}{\sqrt{3}},
\end{equation}
and $H$-type
\begin{equation}
\ket{H} = \cos\Big( \frac{\pi}{8}\Big)\ket{0} + \sin\Big(\frac{\pi}{8}\Big)\ket{1}.
\end{equation}
In order to quantify the magic in any given state, we use the stabiliser 2-R{\'e}nyi entropy \cite{leone2022stabilizer}. For a single qubit pure state, $\ket{\psi}$, the stabiliser 2-R{\'e}nyi entropy is defined as 

\begin{equation}
M_2(\ket{\psi})  = -\log_2\Bigg( \frac{1}{4}\sum_{P\in \mathcal{P}_1}\bra{\psi}P\ket{\psi}^4\Bigg) - \log_2(2),
\end{equation}
where $\mathcal{P}_1 = \{ \mathbb{I},X,Y,Z\}$. We also define the stabiliser 2-R{\'e}nyi entropy for a density matrix as
\begin{equation}\label{eq:magicdensitymat}
M_2(\rho) = -\log_2\Big(\frac{1}{4}\sum_{P\in \mathcal{P}_1}\big(\Tr\big(P\rho \big)\big)^4 \Big) - \log_2(2).
\end{equation}

\subsection{GNU codes}
The permutation-invariant codes that we will look at are the gnu codes \cite{ouyang2014permutation}. They are defined by their logical states 
\begin{equation}\label{eq:gnustates}
\begin{split}
\ket{0_{g,n,u}} &= \sqrt{2^{-(n-1)}}\sum_{\substack{0\leq j\leq n\\j \;\text{even}}} \sqrt{\begin{pmatrix}
    n\\j
\end{pmatrix}}
\ket{D^N_{gj}},\\
\ket{1_{g,n,u}} &= \sqrt{2^{-(n-1)}}\sum_{\substack{0\leq j\leq n\\j \;\text{odd}}} \sqrt{\begin{pmatrix}
    n\\j
\end{pmatrix}}
\ket{D^N_{gj}},
\end{split}
\end{equation}
where the number of qubits $N = gnu$ and the Dicke states are defined as
\begin{equation}
\ket{D^N_w} = \frac{1}{\sqrt{\binom{N}{w}}}\sum_{\substack{x_1, ..., x_N \in \{0,1\} \\x_1+...+x_N = w}}\ket{x_1,...,x_N}.
\end{equation}

\section{Derivation of distillation protocol}\label{sec:distillationderivation}
The state $\rho_{N,\epsilon}$ is written in terms of the tensor product state $\ket{\phi_{\bf x}}$, in which we group together all the states of the same weight and take the inner product with a Dicke state of weight $s$ to get
\begin{equation}\label{eq:innerprod}
\langle D_s^N|\phi_{\textbf{x}}\rangle = \frac{1}{\sqrt{\binom{N}{s}}} \sum_{t=\max(0,s-(N-\omega))}^{\min(s,\omega)} \alpha_{s,t} \binom{\omega}{t}\binom{N-\omega}{s-t}.
\end{equation}
where $|\textbf{x}| = \omega$, $N=gnu$ and 
\begin{equation}
\begin{split}
\alpha_{s,t} &= (\cos v)^{N-\omega-(s-t)}(e^{i\theta}\sin v)^{s-t}(\sin v)^{\omega-t}(-e^{i\theta}\cos v)^t\\
& = (-1)^te^{is\theta}(\cos v)^{N-(s+\omega-2t)}(\sin v)^{s+\omega-2t}\\
& = (-1)^te^{is\theta}(\cos v)^N(\tan v)^{s+\omega-2t}
\end{split}
\end{equation}
which, for ease of notation, we have defined $\alpha_{s,t}\equiv \alpha_{s,t}(g,n,u,v,\theta,\omega)$.
We will apply the projection 
\begin{equation}
\Pi = \ket{0_{g,n,u}}\bra{0_{g,n,u}}+\ket{1_{g,n,u}}\bra{1_{g,n,u}}
\end{equation}
so that the resulting $N$-qubit state, after normalising, is of the form
\begin{equation}\label{eq:rhofinal}
\rho_{\text{final},\epsilon} =\frac{\Pi \rho_{N,\epsilon} \Pi}{\Tr[\Pi \rho_{N,\epsilon} \Pi]}  = \frac{1}{a+b}\begin{pmatrix} a & c\\
c^* & b
\end{pmatrix} .
\end{equation}
We calculate
\begin{equation}
\begin{split}
&\langle 0_{g,n,u}|\phi_{\textbf{x}}\rangle = \\
&\sqrt{2^{-(n-1)}}\sum_{\substack{0\leq j \leq n\\j\;\text{even}}}\frac{\sqrt{\binom{n}{j}}}{\sqrt{\binom{N}{gj}}}\sum_{t=\max(0,gj-(N-\omega))}^{\min(gj,\omega)}\alpha_{gj,t}\binom{\omega}{t}\binom{N-\omega}{gj-t}
\end{split}
\end{equation}
and similarly
\begin{equation}
\begin{split}
&\langle 1_{g,n,u}|\phi_{\textbf{x}}\rangle =\\
&\sqrt{2^{-(n-1)}}\sum_{\substack{0\leq j \leq n\\j\;\text{odd}}}\frac{\sqrt{\binom{n}{j}}}{\sqrt{\binom{N}{gj}}}
\sum_{t=\max(0,gj-(N-\omega))}^{\min(gj,\omega)}\alpha_{gj,t} \binom{\omega}{t}\binom{N-\omega}{gj-t}.
\end{split}
\end{equation}
Let us define 
\begin{equation}
\beta_j(g,n,u,v,\theta,\omega) = \frac{\sqrt{\binom{n}{j}}}{\sqrt{\binom{N}{gj}}} \sum_{t=\max(0,gj-(N-\omega))}^{\min(gj,\omega)}\alpha_{gj,t} \binom{\omega}{t}\binom{N-\omega}{gj-t},
\end{equation}
so that we can now write
\begin{equation}\label{eq:generalabc}
\begin{split}
a &= 2^{-(n-1)} \sum_{\omega=0}^N \binom{N}{\omega}\epsilon^{\omega}(1-\epsilon)^{N-\omega} \Big| \sum_{\substack{0\leq j \leq n\\j\; \text{even}}} \beta_j\Big|^2 \\
b &= 2^{-(n-1)} \sum_{\omega=0}^N \binom{N}{\omega}\epsilon^{\omega}(1-\epsilon)^{N-\omega} \Big| \sum_{\substack{0\leq j \leq n\\j\; \text{odd}}} \beta_j\Big|^2  \\
c &= 2^{-(n-1)} \sum_{\omega=0}^N \binom{N}{\omega}\epsilon^{\omega}(1-\epsilon)^{N-\omega} \Big( \sum_{\substack{0\leq j \leq n\\j\; \text{even}}} \beta_j\Big)\Big( \sum_{\substack{0\leq j \leq n\\j\; \text{odd}}} \beta_j\Big)^*
\end{split}    
\end{equation}
where, for simplicity he have defined $\beta_j\equiv \beta_j(g,n,u,v,\theta,\omega)$.
We then perform a decoding strategy \cite{ouyang2021permutation} to bring us from the logical states $\ket{0_{g,n,u}}$ and $\ket{1_{g,n,u}}$ to the single qubit states $\ket{0}$ and $\ket{1}$ respectively.

\section{Final states for $g=n=1$ and $u=2,3,4$}\label{sec:appfinalstates}
The final state, written in the form of eq. \eqref{eq:finalstate} for $g=n=1$ and $u=2$ is
\begin{equation}\label{eq:abc2}
\begin{split}
a &= - \frac{1}{4} \Big(1 + (1 - 2 \epsilon) \cos(2 v)\Big)^2,\\
b &= \frac{1}{4} - \frac{1}{4} (1 - 2 \epsilon)^2 \cos(4 v),\\
c &= \frac{1}{2\sqrt{2}} e^{-i \theta} (1 - 
 2 \epsilon) \Big(1 +(1-2\epsilon)\cos(2 v)\Big)\sin(2v),
\end{split}
\end{equation}
for $g=n=1$ and $u=3$ it is
\begin{equation}\label{eq:abc3}
\begin{split}
a &= \frac{1}{8} \Big(1 + (1 - 2 \epsilon) \cos(2 v)\Big)^3,\\
b &= \frac{1}{32}\Big( 6-8\epsilon(1-\epsilon) +(3-2\epsilon)(1-4\epsilon^2)\cos(2v)\\
&\;\;\;\;\;- 6 (1 - 2 \epsilon)^2 \cos(4v) + 3 (-1 + 2 \epsilon)^3 \cos(6v) \Big),\\
c &= \frac{\sqrt{3}}{8}e^{-i\theta}(1-2\epsilon)\Big( 1-(1-2\epsilon)\cos(2v) \Big)^2\sin(2v).
\end{split}
\end{equation}
and for $g=n=1$ and $u=4$ the final state is of the form
\begin{equation}\label{eq:abc4}
\begin{split}
a &= \Big( -\frac{1}{2} - (\frac{1}{2}-\epsilon)\cos(2v) \Big)^4 \\
b &= -\frac{1}{8}\Big(1+(1-2\epsilon)\cos(2v) \Big)^2 \\
& \;\;\;\;\; \times \Big( -1+2(1-\epsilon)\epsilon+(1-2\epsilon)^2\cos(4v) \Big)\\
c &= \frac{1}{8}e^{-i\theta}(-1+2\epsilon)\Big( -1-(1-2\epsilon)\cos(2v)\Big)^3\sin(2v).
\end{split}
\end{equation}

\section{Bravyi and Kitaev magic state distillation}\label{sec:appBK}
Bravyi and Kitaev \cite{bravyi2005universal} take five imperfect copies of the state $\ket{T}$ such that the initial state has error $\epsilon$, and distil a single state $\ket{T}$ with error suppression 
\begin{equation}\label{eq:paperTerror}
\epsilon_{\text{out}}(\ket{T}) = \frac{t^5 + 5t^2}{1+5t^2+5t^3+t^5}, \;\;\; t = \frac{\epsilon}{1-\epsilon}
\end{equation}
and probability of success
\begin{equation}\label{eq:paperTps}
p_s (\ket{T}) = \frac{\epsilon^5+5\epsilon^2(1-\epsilon)^3+5\epsilon^3(1-\epsilon)^2+(1-\epsilon)^5}{6}.
\end{equation}
To distil the state $\ket{H}$, they use a 15-qubit code to achieve
\begin{equation}\label{eq:paperHerror}
\epsilon_{\text{out}}(\ket{H}) = \frac{1-15(1-2\epsilon)^7+15(1-2\epsilon)^8-(1-2\epsilon)^{15}}{2\Big( 1+12(1-2\epsilon)^8 \Big)},
\end{equation}
and 
\begin{equation}\label{eq:paperHps}
p_s (\ket{H}) = \frac{1+15(1-2\epsilon)^8}{16}.
\end{equation}

\section{2-Qubit repetition code comparison}\label{sec:2rep}

\begin{figure*}[ht]
\centering
 \includegraphics[width=0.5\linewidth]{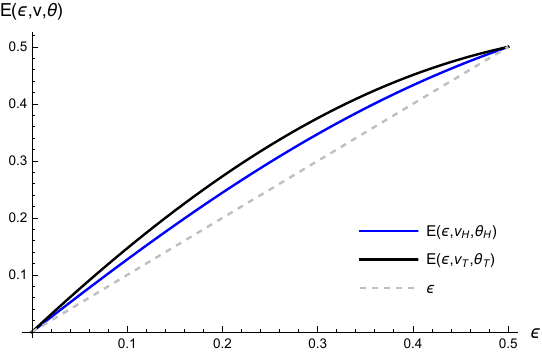}%
\caption{Maximum error (see Eq. \eqref{eq:maxerror}) for the 2-qubit repetition code, a special case of the gnu code with $g=2$ and $n=u=1$. The error is shown for initial states defined by the parameters given in Eq. \eqref{eq:Trep} (black curve) and Eq. \eqref{eq:Hrep} (blue curve). These initial states (see Eqs. \eqref{eq:initialstates} and \eqref{eq:initial}) enable the distillation of $\ket{T}$ and $\ket{H}$ respectively, but provide no error suppression.}
\label{Fig4}
\end{figure*}

The 2-qubit repetition code is a special case of the gnu code with $g=2$ and $n=u=1$. For the initial states defined in Eqs. \eqref{eq:initialstates} and \eqref{eq:initial}, choosing the parameters
\begin{equation}\label{eq:Trep}
v_T = \arcsin\Big( \sqrt{(1+\sqrt{2}- \sqrt{3})/2}\Big), \;\;\;\;\; \theta_T = 7\pi/8, 
\end{equation}
and using eqs. \eqref{eq:rhofinal} and \eqref{eq:generalabc}
allows the state $\rho_{\text{final},\epsilon=0} = \ket{T}\bra{T}$ to be distilled, assuming there is no noise. Similarly, choosing
\begin{equation}\label{eq:Hrep}
v_H = \text{arccot}\Big( \sqrt{1+\sqrt{2}} \Big), \;\;\;\;\; \theta_H = 0,    
\end{equation}
distils the state $\rho_{\text{final},\epsilon=0} = \ket{H}\bra{H}$. However, in the presence of noise, figure \ref{Fig4} shows that the maximum error (see eq. \eqref{eq:maxerror}) is greater or equal to the input error. Therefore, the 2-qubit repetition code does not provide any error suppression when distilling the magic states $\ket{T}$ or $\ket{H}$.  

\end{document}